\documentstyle[12pt,aasms4]{article}
\def\day{{}$^{\rm d}$\llap{.}}

\begin{document}

\title{THE RR LYRAE PERIOD-AMPLITUDE RELATION AS A CLUE TO THE ORIGIN OF
THE OOSTERHOFF DICHOTOMY}

\author{Christine M. Clement}

\affil{Department of Astronomy, University of Toronto \\
Toronto, Ontario, M5S 3H8, CANADA\\
electronic mail: cclement@astro.utoronto.ca}

\author{Ian Shelton}

\affil{Subaru Telescope, National Astronomical Observatory of Japan\\
650 North A'ohoku Place, Hilo, Hawaii, 96720 USA \\
electronic mail: shelton@subaru.naoj.org}

\begin{abstract}
An examination of the period-$V$ amplitude relation for RRab stars 
(fundamental mode pulsators) with `normal' light curves in the
Oosterhoff type I clusters M3 and M107 and in the Oosterhoff type II
clusters M9 and M68 reveals that the $V$ amplitude for a given period
is not a function of metal abundance. Rather, it is a function of the 
Oosterhoff type. This result is confirmed by published observations of
RRab stars in M4, M5 and M92. A method devised by Jurcsik and Kov\'acs has been
used to determine whether the light curve of an RRab star is `normal' or
`peculiar'. Although M3 is considered to belong to the Oosterhoff type I
group, it has three bright RRab stars that seem to fit the period-amplitude
relation for Oosterhoff type II RRab stars. There is evidence that
these bright stars are in
a more advanced evolutionary state than the other RRab stars in M3, thus 
leading to the conclusion that the Oosterhoff dichotomy is due to evolution. 
Our result gives support to the Lee, Demarque \& Zinn hypothesis
that most RR Lyrae variables in Oosterhoff type I clusters are ZAHB objects
while those in the Oosterhoff type II clusters are more evolved.
This may have important implications for the derived ages of
Oosterhoff type II clusters. If their RR Lyrae variables
have all evolved away from the ZAHB, then their ages have been overestimated 
in studies that assume they are ZAHB objects.

\end{abstract} 
\keywords{
globular clusters: individual (M3, M4, M5, M9, M68, M92, M107) ---
stars: fundamental parameters ---
stars: horizontal-branch ---
stars: variables: RR Lyrae 
}
%
%
\section{INTRODUCTION}
It has often been assumed that the period-amplitude relation for
fundamental mode RR Lyrae (RRab) stars is a function of metal abundance. 
Forty years ago, in a study of the period-$m_{pg}$ amplitude
diagram for approximately 50 field RRab
stars, Preston (1959) demonstrated that 
the more metal-poor and more metal-rich stars stars appeared to define two
sequences well separated in amplitude. 
Subsequently, Dickens \& Saunders (1965) and 
Dickens (1970) found similar results for RRab stars in globular clusters.
Then, in a study of six well observed clusters, Sandage (1981b, hereafter 
referred to as S81b) quantified
this result. He measured a period shift, 
$\Delta \log P$, for each cluster relative to the
mean period-amplitude relation for M3 and found a correlation between
$\Delta \log P$ and metal abundance in the sense that the more metal poor
RR Lyrae variables had longer periods. As a result of this,  
some investigators have used the
period-amplitude relation as an indicator of metal abundance, particularly
in faint systems where it is difficult to estimate it by other methods. 
Sandages's (S81b) result was based mainly on photographic data, but
in the last ten years, CCD detectors have been widely
used for observations of RR Lyrae variables in globular clusters. 
CCDs are linear detectors and this makes the photometry more accurate.
They also have a higher quantum efficiency than
photographic emulsions so that exposure
times can be shorter. As a result, the time and magnitude at maximum and
minimum light can be more precisely established and it is 
possible to derive more accurate amplitudes. 
Another problem with earlier studies was that, in some cases, stars with
non-repeating light curves, stars that exhibit the Blazhko effect,
were included in the samples. The amplitude of light variation for
Blazhko stars varies over time scales longer than the basic pulsation
period. 
Thus, if Blazhko variables are not identified 
they introduce
scatter into the period-amplitude (P-A) diagram. To address this problem,
Jurcsik \& Kov\'acs (1996, hereafter JK) recently
devised a compatibility test for identifying Blazhko variables.

The purpose of our investigation is 
to use $V$ amplitudes derived from CCD photometry
to re-examine the P-A relation 
for RR Lyrae variables in globular clusters of both Oosterhoff types
and to apply
JK's test so that Blazhko variables can be identified.
In Table 1, we list [Fe/H], horizontal branch  classification and Oosterhoff
(1939, 1944) type for
seven clusters for which published $V$ photometry is
available.
The [Fe/H] is on the system of Jurcsik (1995, hereafter J95)
and the horizontal branch (HB) classification is indicated by the quantity
(B-R)/(B+V+R) defined by Lee, Demarque \& Zinn (1994). 
A negative value for this quantity indicates that most of
the HB stars are on the red side of the instability strip and a high positive
value indicates that most are on the blue side. 
%
%

\section{THE COMPATIBILITY TEST OF JURCSIK \& KOV\'ACS}

The modus operandi of JK was to
characterize the light curve systematics of a sample of  74 RRab stars
with normal light curves by studying the interrelations among the Fourier 
parameters. Specifically, they derived a set of 9 equations for calculating 
the Fourier amplitudes $A_1$ to $A_5$ and the phase differences $\phi_{21}$
to $\phi_{51}$. If, for a particular star, the calculated value for any one
of the parameters is not in good agreement with the 
observed value, the star's
light curve is considered to be peculiar. They illustrated the effectiveness
of their method with a study of RV UMa, an RR Lyrae star that exhibits the 
Blazhko effect. 
An independent demonstration of the validity of JK's test comes from
a study of V12
in NGC 6171 (M107) by Clement \& Shelton (1997, hereafter CS97). CS97
observed this star in two different
years and the light curve appears to repeat well 
from cycle to cycle. Nevertheless, the JK compatibility
test indicates that the light curve of V12 is peculiar. 
It turns out that this is indeed the case if one compares
the light curve of CS97 with an earlier one published by Dickens (1970).
Both the shape and amplitude of the
curve of V12 changed dramatically between the two epochs.

%
%

\section{THE PERIOD-AMPLITUDE RELATION OF RR LYRAE VARIABLES}

\subsection{The Oosterhoff type I clusters M3 and M107}

In the two upper panels of Figure 1, 
we show the period-$V$ amplitude relations for the
Oosterhoff type I (OoI) globular clusters, M3 and M107. The data  for M3
are from 
Kaluzny \it et al. \rm (1998, hereafter KHCR) and for M107
from CS97.  
To establish which RRab 
stars in M3 and M107 had peculiar light curves, we applied JK's compatibility
test using equations recently derived 
by Kov\'acs \& Kanbur (1998) from a sample of 257 RRab stars.
What we see in
the upper two panels of the figure is that most of the RRab stars with peculiar 
light curves (the open circles) have lower amplitudes than other stars with 
the same period.
According to Szeidl (1988) and JK, the maximum amplitude of a Blazhko variable
fits the period-amplitude relation for regular RRab stars.
Thus these stars are probably Blazhko variables which were not
observed at a time when the amplitude was at its maximum. 
In the M3 plot, the
squares represent three regular RRab stars (V14, V65 and V104) 
that are brighter that the other stars. 
KHCR concluded that these three stars
are probably in a more advanced evolutionary state than the others. 
In the lower panel of Figure 1, we plot the RRab stars,
V29 in M4 and V8 and V28 in M5. These are stars that JK classified 
as normal and for which
published $V$ photometry is available.  Clementini \it et al. \rm (1994) 
observed M4 and  Storm \it et al. \rm (1991) observed M5. 
Also plotted are the $V$ amplitudes that CS97
derived from Reid's (1996) observations of RRc stars in M5.

The straight line shown in each panel
is a least squares fit to the principal sequence of
regular RRab stars in M3 (the solid circles).
We can readily see that the regular RRab stars in the three OoI clusters
(M107, M4 and M5) fit the P-A relation for M3. There is no evidence for a
shift in $\log P$ even though all three
of these clusters are more metal rich than M3. 
The situation may be different, however, 
for the RRc stars. The curve to the left of $\log P= -0.4$ in each
panel of the diagram is a fit to the RRc stars in M3 
and in this case, the P-A
relations for M5 and M107 are shifted to shorter periods than M3.

\subsection{The Oosterhoff type II clusters M9 and M68 }

In the upper panel of Figure 2, we plot the P-$A_V$ diagram for M9, based on
data published by Clement \& Shelton (1999). 
The solid straight line is a least squares fit to RRab stars in M9 and the
dashed line is the fit to M3 shown in Figure 1. The two  curves 
are the fits to the RRc stars in M3 and M9. 
M3 is among the most metal poor of the OoI clusters and M9 is among the most
metal rich of the OoII clusters, but the diagram shows that
there is a definitive period shift
between the two, for both RRab and RRc stars. This is the Oosterhoff dichotomy. 
In the center panel of Figure 2, the M68 data of Walker (1994) are plotted.
JK found that two RRab stars in M68 (V23 and V35) had normal curves and 
so these are plotted as solid circles. The remaining RRab stars are plotted
as open circles. 
The M68 RRab stars with peculiar light curves generally have lower amplitudes
than those with normal curves, like M3 and M107 in Figure 1. 
However, this trend is not apparent in M9. Perhaps, this is because the stars
with irregular light curves are observed at maximum amplitude.
In the lower panel of Figure 2, we plot amplitudes derived from the 
observations
of Carney \it et al. \rm (1992) for two M92 RRab stars classified
as regular by JK. 
Also included in the lower panel are the three bright stars in M3. 
All of the regular RRab stars plotted in Figure 2 seem to fit one P-A
relation; there is no correlation with
metallicity for the OoII clusters. Also there is no evidence that the 
periods of the RRc stars in M68
are longer than those in M9, even though M68 is more metal poor.

The plots of Figures 1 and 2 demonstrate that the P-A relation for RRab stars
is not a function of metal abundance. Rather, it is related to
Oosterhoff type.
Lee, Demarque \& Zinn (1990, hereafter LDZ) have proposed that
evolution away from the ZAHB plays a role in the Oosterhoff dichotomy.
If this is correct, then the fact that we have found two different P-A 
relations suggests that there may be one P-A relation for
ZAHB stars and another for stars that are more evolved. ZAHB stars have
helium fusion in the core, but after the core helium is consumed, the helium
fusion occurs in a shell. Apparently, this circumstance has more importance 
than metal abundance for determining the P-A relation of fundamental
mode pulsators. 

The situation for the RRc stars (the first overtone pulsators) may be 
different. It is possible that the P-A relation for RRc stars in
OoI clusters depends on metal abundance.

\subsection{The unique case of M3}

For OoI clusters like M3, the models of LDZ
predict that RR Lyrae stars evolve blueward across the instability strip
on the ZAHB, but
when they evolve away from the ZAHB, they become brighter and redder. 
Clement \it et al. \rm (1997) found evidence for blueward evolution of
the M3 star V79 because of a mode switch. Before 1962, V79 was an RRab 
star with a period of 0\day 483, but in 1996,
it was an RRd star with the first overtone mode dominant. This mode switch 
has since been confirmed by Corwin, Carney \& Allen (1999) and Clement \&
Goranskij (1999) who found that it occurred in 1992. 
The mean V magnitude of V79 is $15.71$ comparable to the mean magnitude
($15.69$) of the 21 RRab stars that fit the P-A relation for
OoI clusters (the solid dots in Fig.~1), presumably all ZAHB stars. 
However, the three stars V14, V65 and V104 are brighter. 
Consequently, KHCR concluded that they are 
in a more advanced evolutionary state.
In addition, these three bright stars fit better to the P-A
relation for the OoII RRab stars and their mean 
period is 0\day 625, which is an appropriate value for an OoII cluster. 
In Table 2, we
list their periods, period shifts ($\Delta \log P$), their mean $V$ magnitudes 
and $\Delta V$ relative to other RRab stars with the same amplitude (i.e. the
ones that fit on the straight line of Figure 1). 
The above discussion  indicates that there are
RR Lyrae variables with characteristics of the two Oosterhoff groups in this
one cluster. Assuming
there is no variation in metal abundance among the stars of
M3, this is further evidence that the P-A relation for RRab stars is not a 
function of metal abundance.  

If the P-A relation is not correlated with metal abundance, then why was such
a correlation found by previous investigators?
One reason  is that OoII clusters are, in general, more metal 
poor than OoI clusters. As a consequence of this, a  difference in the 
P-A relation for clusters of the different Oosterhoff groups could be 
attributed to a difference in metal abundance. However, Sandage's
analysis indicated that the $\Delta \log P$-metal abundance 
correlation exists among clusters of one Oosterhoff type. This is documented
in column 11 of Table 7 in his paper (S81b).  The P-A relations
for the OoI clusters  M4 and NGC 6171 (M107) are shifted to short 
periods compared with M3.
We believe that this apparent correlation probably occurs because of a selection
effect in the choice of the M3 sample.
The M3 data were taken from the study of Roberts and Sandage (1955, hereafter
RS) whose
objective was to determine reliable colors for RR Lyrae variables,
and so they excluded stars with non-repeating light curves. 
As a result, stars like those plotted as open circles in
our P-A relation for M3 were not included in their study. This 
makes the P-A relation for M3 appear to be shifted to longer
periods than the other OoI clusters. In addition, the fact that some of the 
M3 RR Lyrae variables have OoII characteristics must also be a contributing
factor. (It must also be acknowledged that in clusters like M4 and M107, the
period at which the transition between fundamental and overtone mode
pulsation occurs is shorter than in M3. However, the short period fundamental
mode pulsators in these clusters have non-repeating light curves.)

M3 is not the only cluster that has RR Lyrae
variables with the characteristics of both Oosterhoff groups.
Omega Centauri is another.
In a study of $\omega$ Centauri, Butler, Dickens \& Epps (1978, hereafter 
BDE) commented that although it
is generally assumed to be an OoII cluster, some of its RR Lyrae
variables have OoI characteristics. This can account for the S81b  
finding that its
P-A relation is shifted to shorter periods than that of M15
and to longer periods than M3.

\subsection{The Period-Luminosity-Amplitude Relation} 

Sandage (1981a, hereafter S81a) showed that there 
is a period-luminosity-amplitude relation for RRab stars in the sense that,
for a given amplitude,
brighter stars have longer periods. He demonstrated
this with photographic observations of two clusters,
M3 (RS) and $\omega$ Centauri (BDE). 
His approach was to calculate the mean apparent bolometric magnitude for
the RR Lyrae variables in a particular cluster and then use van Albada 
\& Baker's (1971, hereafter VAB) equation relating pulsation period, mass, 
temperature and absolute
bolometric magnitude to derive a `reduced' period for each star.  
This is the period the star 
would have if its $m_{bol}$ had the same value as the
cluster mean. 
A plot of amplitude against `reduced' period shows much less scatter than
a regular period-amplitude plot, and if
the masses of the RR Lyrae stars in a particular cluster are the same, this 
implies that the amplitude is a function of temperature.
S81b's correlation between $\Delta \log P$ and
metal abundance was derived from P-A relations plotted with `reduced'
periods. Nevertheless, the plot for NGC 6171 shows considerable scatter. We
believe that this
can be attributed to the fact that Blazhko variables were included in
his sample. The P-L-A relation does not apply to Blazhko variables. 

It is significant that Sandage showed that the P-L-A relation holds for
both $\omega$ Centauri and M3 because these two clusters have
stars with properties of both Oosterhoff groups. 
Even though the P-A
relation for RRab stars is a function of Oosterhoff type, a P-L-A relation 
seems to be
valid for stars in a cluster that belongs to both groups. If there is a 
unique period-amplitude relation for RRab stars on the ZAHB, then the
P-L-A relation can be used to estimate the apparent magnitude of the
ZAHB in any cluster, regardless of its Oosterhoff type, as long as it has 
RRab stars with normal light curves. However,
an examination of the data in Table 2 indicates that the period shift
(at constant amplitude) for the three bright stars in M3 can not be
completely accounted for by a difference in luminosity.
According to VAB's pulsation equation, $\Delta \log P/\Delta m_{bol}$ is $0.34$
for constant mass and temperature, but the
mean $\Delta \log P/\Delta V$ derived for the three bright stars in M3
is $0.48$ with $\sigma=0.10$. Thus there must be a difference in mass and/or
temperature as well.

%
%

\section{IMPLICATIONS FOR AGES OF GLOBULAR CLUSTERS}

In recent years, there has been condsiderable discussion in the literature
(e.g. Chaboyer, Demarque \& Sarajedini 1996; Stetson, Vandenberg \& Bolte 1996) 
about the range of ages of galactic globular clusters and also 
about the question of whether or
not metal poor clusters are older than metal rich clusters.  These issues
have not yet been resolved, but our
results may have some impact on this problem.  In some
investigations, e.g. Gratton (1985), the cluster age is derived from the 
difference between the
ZAHB and the main sequence turnoff (${\Delta V}^{ZAHB}_{TO}$). In these cases, 
the faintest stars on the HB in the vicinity of the
RR Lyrae instability strip are assumed to be ZAHB objects, but if they are in 
fact, at a more advanced evolutionary state, the apparent 
luminosity of the ZAHB is overestimated. As a result, ${\Delta V}^{ZAHB}_{TO}$ 
is also overestimated and this leads to overestimation of
the cluster age. This could cause an apparent age-metallicty relation.

It will be very interesting to see if other studies of RR Lyrae variables
confirm our conclusion that the period-amplitude relation for fundamental
mode pulsators depends on evolutionary state and not on metal abundance.

%
%
\acknowledgements

We would like to thank Pierre Demarque and Norman Simon 
for discussing this work during the course of the investigation. 
They both encouraged us to try to understand the significance of our results 
and we hope we have succeeded.
Thanks are also due to Jason Rowe for his assistance in the preparation of the
diagrams. The work has been
supported by the Natural Sciences and Engineering Research Council of Canada.

\newpage


\begin{figure}
\plotfiddle{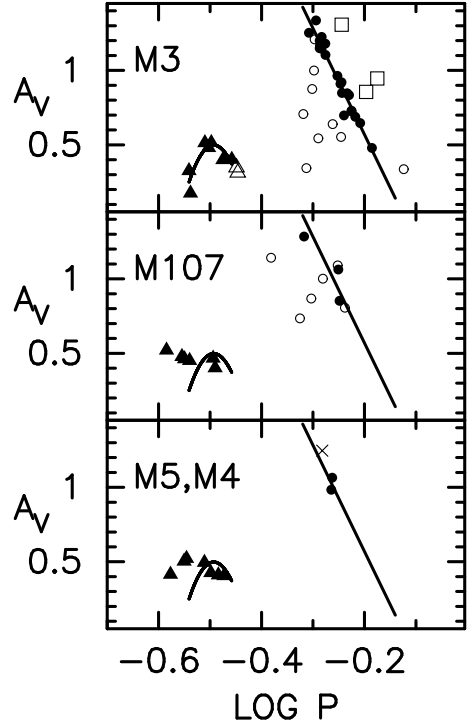}{4.7truein}{0}{80}{80}{-200}{-70}
\caption
{The period-amplitude 
relation ($A_V$ versus $\log P$) for the RR Lyrae
variables in OoI clusters.  
In the two upper panels,
the symbols are as follows: solid triangles for the
RRc stars, open triangles for the RRd stars (M3)
plotted with their overtone
periods, open circles for the RRab stars with peculiar light curves according
to JK's compatibility test and solid circles for the other RRab stars.
The open  squares in the M3 plot are three RRab stars that are in a more
advanced evolutionary state than the others.
In the lower panel, the triangles represent the RRc stars in M5, the solid
circles represent two M5 RRab stars that JK classified as regular and the
cross represents an M4 RRab star that they classified as regular.
In all three panels, the
curve at $\log P$ less than $\sim -0.4$ represents the fit to the 
RRc stars in M3 and the straight line was
derived from a least squares fit 
to the principal sequence of regular RRab stars in M3 (the solid circles).}
\end{figure} 

\newpage
\begin{figure}
\plotfiddle{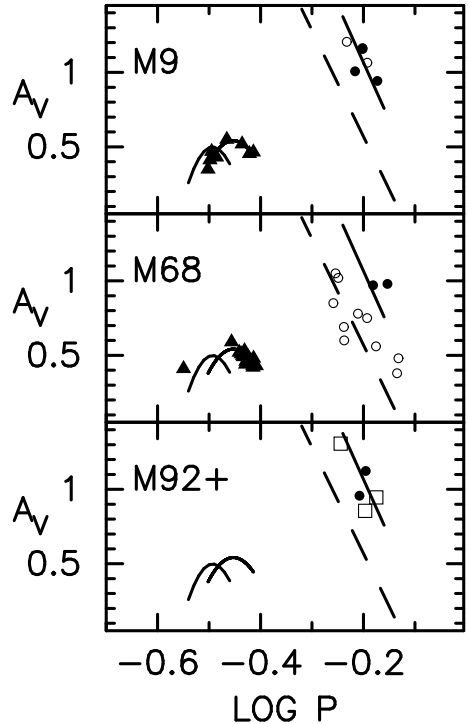}{4.7truein}{0}{80}{80}{-200}{-70}
\caption
{The period-amplitude relation for the OoII clusters. In the two upper 
panels, the symbols are: solid triangles for the RRc stars,
solid circles for regular RRab stars and open circles for the ones
with peculiar light curves. In the lower panel, the solid circles represent  
two regular RRab stars in the OoII cluster  M92 and the three squares are
the three evolved stars in the OoI cluster M3. 
In each panel, the
solid straight line is a least squares fit to RRab stars in M9 and the
dashed line is the fit to M3 shown in Figure 1. The two  curves 
are the fits to the RRc stars in M3 and M9. The
curve and line for M9 are both displaced to longer periods than those for
M3, demonstrating that the RR Lyrae variables in OoII clusters have
longer periods than those in the OoI clusters.
The diagram shows that metal abundance does not affect the P-A relation
for `regular' RRab stars in these three OoII  clusters 
and that the three bright stars in M3 lie close to the OoII P-A relation.} 
\end{figure}

\clearpage

%

%
\newpage
\begin{table}
\caption{Cluster Metal Abundances and HB Type \label{Table 1}}
\begin{flushleft}
\begin{tabular}{lccc}
\tableline
\tableline
Cluster & [Fe/H] & (B-R)/(B+V+R) & Oosterhoff \cr
        &  (J95) &  (LDZ94)     &  type  \cr
\tableline
M107 & -0.68  & -0.76 & I  \cr
M4   & -1.11  & -0.07 & I  \cr 
M5   & -1.25  &  0.19 & I  \cr
M3   & -1.47  &  0.08 & I  \cr
M9   & -1.72  &  0.87 & II \cr
M68  & -2.03  &  0.44 & II \cr
M92  & -2.31  &  0.88 & II \cr
\tableline
\end{tabular}
\end{flushleft}
\end{table}

%
\begin{table}
\caption{The Bright RRab Stars in M3  \label {Table 2}}
\begin{flushleft}
\begin{tabular}{lcccc}
\tableline
\tableline
Star & Period & $\Delta \log P$ & $<V>$ & $\Delta V$  \cr
     & (days) &                 &       &            \cr
\tableline
V14  & 0.6359019 & 0.044        & 15.57 & 0.13 \cr
V65  & 0.6683397 & 0.078        & 15.53 & 0.15 \cr
V104 & 0.5699231 & 0.059        & 15.60 & 0.10 \cr
\tableline
\end{tabular}
\end{flushleft}
\end{table}

\end{document}